\documentclass[aps,prb,twocolumn,superscriptaddress,floatfix,nofootinbib]{revtex4-2}
\usepackage{epsfig,amsmath,amssymb,color,comment,physics}
\usepackage[makeroom]{cancel}
\usepackage[caption=false]{subfig}
\usepackage{mathrsfs}
\usepackage[countmax]{subfloat}
\usepackage[normalem]{ulem}
\usepackage[english]{babel}
\usepackage{dsfont}
\usepackage{float}
\usepackage[bookmarks=true,colorlinks,linkcolor=black,urlcolor=NavyBlue,citecolor=RoyalBlue]{hyperref}
\usepackage[dvipsnames]{xcolor}
\usepackage{soul} 
\usepackage{color, xcolor} 

\usepackage{color}

\graphicspath{{./Figures/}} 

\begin{document}

\title{Emergent Anomalous Hydrodynamics at Infinite Temperature in a Long-Range XXZ Model}

\author{Ang Yang} 
\affiliation{School of Physics, Zhejiang Key Laboratory of Micro-nano Quantum Chips and Quantum Control, Zhejiang University, Hangzhou $310027$, China}

\author{Jinlou Ma} 
\affiliation{School of Physics, Zhejiang Key Laboratory of Micro-nano Quantum Chips and Quantum Control, Zhejiang University, Hangzhou $310027$, China}

\author{Lei Ying} 
\email{leiying@zju.edu.cn}
\affiliation{School of Physics, Zhejiang Key Laboratory of Micro-nano Quantum Chips and Quantum Control, Zhejiang University, Hangzhou $310027$, China}

\begin{abstract}
The conventional wisdom suggests that transports of conserved quantities in non-integrable quantum many-body systems at high temperatures are diffusive. However, we discover a counterexample of this paradigm by uncovering anomalous hydrodynamics in a spin-1/2 XXZ chain with power-law couplings. This model, classified as non-integrable due to its Wigner-Dyson level-spacing statistics in the random matrix theory, exhibits a surprising superdiffusive-ballistic-superdiffusive transport transition by varying the power-law exponent of couplings for a fixed anisotropy.
Our findings are verified by multiple observables, including the spin-spin autocorrelator, mean-square displacement, and spin conductivity. Interestingly, we further quantify the degree of quantum chaos using the Kullback-Leibler divergence between the entanglement entropy distributions of the model's eigenstates and a random state. Remarkably, an observed local maximum in the divergence near the transition boundary suggests a link between anomalous hydrodynamics and a suppression of quantum chaos. This work offers another deep understanding of emergent anomalous transport phenomena in a wider range of non-integrable quantum many-body systems.
\end{abstract}

\date{\today}
\maketitle

\section{Introduction}

Recently, probing the emergent hydrodynamics with a few conserved quantities in quantum many-body systems has attracted wide interest in both experimental and theoretical communities~\cite{joshi2022observing,Schuckert2020,wei2022quantum,Ljubotina2019KPZ,Dupont2020universal,Steinigeweg2015typical,Bingtian2020,Lux2014tail,bulchandani2021superdiffusion,bush2020hydrodynamic,Bertini2021review,gopalakrishnan2022anomalous}. In general, ballistic transport is considered to mostly occur in integrable quantum many-body systems despite some counterexamples. These systems have stable quasiparticles that propagate without scattering, leading to ballistic transport, while the standard diffusion is expected for non-integrable systems~\cite{gopalakrishnan2022anomalous}. 
However, recent works have revealed that both integrable and non-integrable models can exhibit unusual late-time transports with different mechanisms~\cite{Feldmeier2020anoma,McRoberts2022classicalanomal,Menu2020anomalous,Richter2022anoma,Zotos1999xxz,Richter2019xxz,Kloss2019longrange,Kloss2020,Lux2014tail,gopalakrishnan2022anomalous,Brenes2018impurity}. 
On the experimental side, associated with the great development of various quantum simulators, in particular such as cold atoms in optical lattice~\cite{gross2017atom,Hild2014atom,jepsen2020atom,schafer2020atom,zhang2017atom,wei2022quantum}, Rydberg atoms~\cite{Andersonrydberg,browaeys2020ryberg,geier2021rydberg,Guardado2018rydberg,Hollerith2022rydberg,labuhn2016rydberg,weimer2010rydberg}, superconducting processor~\cite{shi2023probing} and ion trap~\cite{joshi2022observing,blatt2012ion,islam2011ion,lanyon2011ion,schneider2012ion}, a broad class of transport behavior has been observed. 

Among them, the Heisenberg spin system is an important platform to study various quantum transport behaviors at infinite-temperature, such as in 1D spin-1/2 chains~\cite{Schuckert2020,Kloss2019longrange,Zotos1999xxz,Steinigeweg2015typical,Steinigeweg2017typicalbroaden}, ladders~\cite{Richter2019ladderdiff,Marko2013ladder} and 2D lattices~\cite{wei2022quantum}. 
The coupling range and type of spin exchange can govern integrability and transport in quantum many-body systems. For example, in the paradigmatic spin-1/2 XXZ chain, the isotropic point $\Delta=1.0$ exhibits Kardar-Parisi-Zhang (KPZ) superdiffusion. Standard diffusion recovers at $\Delta>1.0$, while ballistic transport prevails for $\Delta<1.0$, as expected for integrable models~\cite{bulchandani2021superdiffusion,Zotos1999xxz,wei2022quantum,Richter2019xxz,Steinigeweg2015typical,Steinigeweg2017typicalbroaden,Bertini2021review,gopalakrishnan2022anomalous}.
On the other hand, long-range interactions are ubiquitous in atomic and ion trapping systems, encompassing a diverse range of types~\cite{Aikawa2021longrange,Lu2012longrange,schauss2012longrange,Saffman2010longrange,yan2013longrange}, and many recent works have focused on the properties of these interactions in long-range models, such as localization~\cite{Burin2015localization,Safavi-Naini2019localization,Schiffer2019localization}, bound states~\cite{kranzl2022bound,Macr2021bound}, nonlocality and entanglement~\cite{Birnkammer2022nonlocal,Cevolani2018nonlocal,Foss-Feig2015nonlocal,Hauke2013nonlocal,Lerose2019nonlocal,Maghrebi2016nonlocal,Ren2020entanglement,Vanderstraeten2018nonlocal,Naldesi2017entanglement}, etc. Interestingly, the spin-1/2 XY chain with power-law couplings, despite being a non-integrable model, exhibits anomalous hydrodynamics. This behavior corresponds to classical Lévy flights, distinguishing from the traditional diffusion
~\cite{joshi2022observing,Schuckert2020,Zaburdaev2015Levy,wei2022quantum,bulchandani2021superdiffusion}. 
Traditionally, studies of spin models have investigated the independent effects of anisotropy and long-range couplings, both of which can induce anomalous transport behaviors. However, these anomalous transports arise from distinct mechanisms and monotonically speed up or slow down as relevant parameters change. Hence, a key question remains: What kind of transport will emerge from the interplay between anisotropy and long-range couplings? One intriguing possibility is the appearance of non-monotonic transitions. Previous work in Ref.~\cite{Kloss2020} explored spin transport in a long-range isotropic Heisenberg model with disorders, finding subdiffusion. However, the effects of anisotropy and long-range couplings were suppressed due to the dominant disorder. More importantly, what signatures characterize these anomalous hydrodynamics, in particular in non-integrable models?

To address these questions, in this paper, we utilize the spin-spin autocorrelation function to directly investigate the magnetization dynamics across the entire Hilbert space at infinite temperature. In a long-range clean spin-1/2 XXZ model, this approach allows us to clearly study distinct transport regimes.
The paper is organized as follows. The model and observables are first introduced in Sec.~\ref{sec:model}. In Sec.~\ref{sec:DQT and TDVP}, we describe in detail the relevant numerical methods, i.e., dynamical quantum typicality~(DQT) and time-dependent variational principle~(TDVP). The numerical results based on different observables are presented and discussed in Sec.~\ref{sec:results}. 

\section{Models and Observables}\label{sec:model}

We investigate the magnetization dynamics of a one-dimensional clean and undriven spin-1/2 XXZ chain with power-law couplings. The system is subjected to a magnetic field applied along $z$-axis~\cite{Bull2022XYZ}. Under open boundary conditions, the Hamiltonian of the system can be written as
\begin{equation}
    \hat{H} = \frac{1}{\mathcal{N}}\sum\limits_{i<j}^L  
    \frac{-J}{{\lvert i-j\rvert}^{\alpha}} \left(\hat{\sigma}_i^{x}\hat{\sigma}_j^{x} + \hat{\sigma}_i^{y}\hat{\sigma}_j^{y} + \Delta\hat{\sigma}_i^{z}\hat{\sigma}_j^{z}\right) + h_z \sum\limits_{i=1}^L \hat{\sigma}_i^z,
\label{hamiltonian}
\end{equation}
where $\hat{\sigma}_i^{\mu}, \mu\in\{x,y,z\}$ are the standard spin-1/2 Pauli matrice of site $i$. $L$, $J$, and $h_z$ denote the system size, the coupling strength, and the magnetic field, respectively. $\Delta>0$ is the ferromagnetic anisotropy along the $z$ direction, and the parameter $\alpha$ determines the strength of the long-range couplings between arbitrary sites in the chain. Here we introduce the Kac norm $\mathcal{N}=\sum\limits_{i<j} {\lvert i-j\rvert}^{-\alpha}/(L-1)$ and it rescales the Hamiltonian to eliminate its dependence on the system size $L$ and the coupling parameter $\alpha$. This ensures that the energy density has the correct scaling behavior, independent of system size~\cite{Bull2022XYZ,Schuckert2020}. 
The anisotropy parameter $\Delta$ breaks a symmetry in the rotating frame. In the isotropic case, $\Delta=1$ with both $U(1)$ and $SU(2)$ symmetries unbroken as well as the total $\vec{\boldsymbol{\sigma}}$ conserved; in the presence of axial anisotropy along $z$-axis, $\Delta\neq 1$ with $U(1)$ symmetry unbroken and the total $\sigma^z$ magnetization conserved. These conserved quantities determine the macroscopic late-time hydrodynamics of the quantum system. In the rest of this paper, we always keep $J=1$ which sets the time unit, as well as magnetic field at a fixed value $h_z=3$~\cite{Bull2022XYZ}. 

\begin{figure}
\centering
\includegraphics[width=0.9\linewidth]{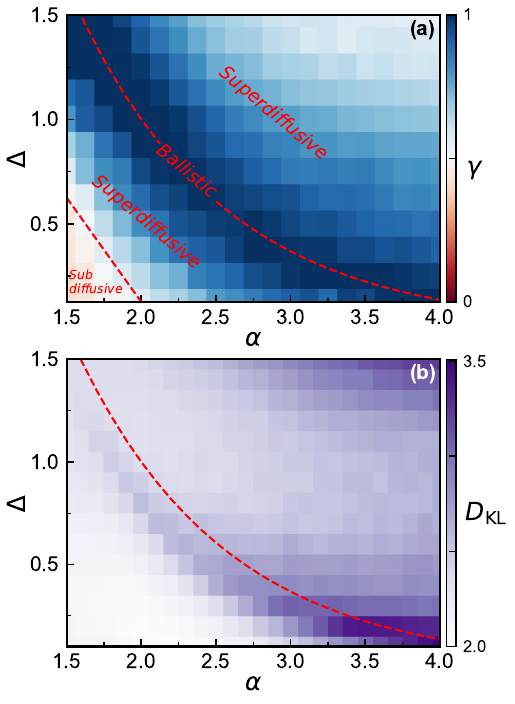}
\caption{
(a) Dynamical exponent $\gamma$ extracted from the autocorrelator $C_0(t)$  as a function of $\Delta$ and $\alpha$. The numerical results are obtained by the DQT method with $L=21$. The ballistic transport (the darkest blue) corresponds to $\gamma\sim1$ and the red region denotes subdiffusion ($0<\gamma<1/2$), while superdiffusion ($1/2<\gamma<1$) dominates in the light blue region. The red dashed line (lower left) marks the boundary between superdiffusive and subdiffusive regimes. (b) The logarithm of Kullback-Leibler (KL) divergence between the microcanonical distribution of eigenstate EE and the Bianchi-Dona (BD) distribution. The system size for panel (b) is $L=14$ with the half-filling condition. The red dashed curve in (a) and (b) is the exponential function $\Delta\sim e^{-\alpha+2}$.
}
\label{phase}
\end{figure}

In general, the transport property can be directly measured by observing how a single spin excitation scatters against a background at an infinite temperature~\cite{joshi2022observing,Schuckert2020}. 
In detail, we initiate a spin excitation at the center of the chain at time $t=0$ by applying $\hat{\sigma}_0^z$.  Then, we track its propagation through space and time by obtaining the unequal-time correlation function at infinite temperature~\cite{joshi2022observing}, which is given by
\begin{equation}
    C_j(t) = \left\langle \hat{\sigma}_j^z(t) \hat{\sigma}_{0}^z(0)\right\rangle_{T=\infty},
\label{correlation}
\end{equation}
where $\hat{\sigma}_j^z(t)=e^{i\hat{H}t}\hat{\sigma}_j^ze^{-i\hat{H}t}$ is the time-dependent operator in the Heisenberg picture with an initial condition of $C_0(0)=1$. 
In particular, the autocorrelator $C_0(t)$ determines the remaining excitation at the central site and it is expected to exhibit anomalous hydrodynamics, as illustrated in Fig.~\ref{phase}.  We will explicitly discuss our results in Sec.~\ref{sec:results}.

In an isolated quantum many-body system, preparing a state with infinite temperature is challenging. According to the canonical thermal ensemble theory at finite temperature, the equilibrium expectation value of an observable is equivalent to the trace of the product of the density operator and the observable~\cite{heitmann2020selected}. Thus, we have 
\begin{equation}
    C_j(t) = \frac{\mathrm{Tr}\left[e^{-\beta \hat{H}}\hat{\sigma}_j^z(t) \hat{\sigma}_{0}^z(0)\right]}{Z},
\label{full trace}
\end{equation}
where $Z=\mathrm{Tr}[e^{-\beta \hat{H}}]$ is the canonical partition function with the reverse temperature $\beta=1/T$. 
Then, we introduce the spatial variance of $C_j(t)$, also known as the mean-square displacement (MSD), is given by~\cite{Kloss2020}
\begin{equation}
    \Sigma^2(t) = \sum\limits_{j=-L/2}^{L/2} j^2C_j(t) - \left(\sum\limits_{j=-L/2}^{L/2}jC_j(t) \right)^2,
\label{MSD}
\end{equation}
which is directly related to the time-dependent diffusion coefficient
$D(t)=d\Sigma^2/2dt$. In the diffusion scenario, $\Sigma(t)\propto t^{1/2}$ and $D(t)$ converge to a constant in the long time limit. Apart from this, the transport is considered as superdiffusive for $\Sigma(t)\propto t^\gamma$ with $\gamma\in (1/2,1)$, ballistic for $\gamma=1$, and subdiffusive for $\gamma\in (0,1/2)$~\cite{Kloss2020,bulchandani2021superdiffusion,Richter2019xxz,Zotos1999xxz,gopalakrishnan2022anomalous,Bertini2021review}. Hence, we define a time-dependent dynamical exponent $\gamma(t)=d\mathrm{log}\Sigma^2(t)/2d\mathrm{log}t$ to extract $\gamma$. In the thermodynamic limit, the spin autocorrelator is expected to decay following a power law~\cite{Dupont2020universal}, expressed as
\begin{equation}
    \lim\limits_{t, L\to\infty}C_0(t)\sim t^{-\gamma}.
\end{equation}

In practice, Eq.~\eqref{full trace} can be expressed as an equally weighted trace over any sets of complete eigenstates, e.g. product states in the $\sigma^z$ basis~\cite{joshi2022observing}. However, it is hard to calculate the trace exactly for a large system as the Hilbert-space dimension is $\mathcal{N}=2^L$. Hence two approximate numerical methods are exploited in Sec.~\ref{sec:DQT and TDVP}.

\section{Numerical approaches}\label{sec:DQT and TDVP}
\subsection{Dynamical quantum typicality}
The dynamical quantum typicality (DQT) indicates that a single pure quantum state can exhibit the same properties of an entire statistical ensemble~\cite{heitmann2020selected,Steinigeweg2015typical,Steinigeweg2017typicalbroaden,jin2021random,Richter2018typical}. It is also effective even in the cases beyond the eigenstate thermalization hypothesis (ETH)~\cite{Srednicki1994ETH,Deutsch1991ETH}, and is a powerful tool in general for the accurate calculation of real-time correlation functions~\cite{heitmann2020selected,Richter2018typical,Steinigeweg2015typical,Elsayed2013powerful,Steinigeweg2014powerful}. 

The main idea amounts to replacing the trace $\mathrm{Tr}\{\bullet\}=\sum_{i} \langle i| \bullet|i \rangle$ in Eq.~\eqref{full trace} by a simple scalar product $\langle \psi|\bullet|\psi\rangle$. Here, $|\psi\rangle$ is a reference pure state, which is randomly drawn from the Hilbert space according to the unitary invariant Haar measure~\cite{Richter2018typical,Bartsch2009typical}. Then, it allows us to rewrite the unequal-time correlation function in Eq.~\eqref{full trace} as~\cite{Steinigeweg2015typical,Elsayed2013powerful,Steinigeweg2014powerful,Steinigeweg2014correlation}
\begin{equation}
    C_j(t) = \frac{\mathrm{Re}\left\langle \psi|e^{-\beta \hat{H}}\hat{\sigma}_j^z(t) \hat{\sigma}_{0}^z(0)|\psi\right\rangle}{\left\langle \psi|e^{-\beta \hat{H}}|\psi\right\rangle} + \epsilon(|\psi\rangle).
\end{equation}
At $T\rightarrow \infty$, the correlation function can be written as
\begin{equation}
    C_j(t) = \frac{\mathrm{Re}\left\langle \psi(t)|\hat{\sigma}_j^z |\phi(t)\right\rangle}{\left\langle \psi(0)|\psi(0)\right\rangle} + \epsilon(|\psi\rangle),
\label{approximation}
\end{equation}
where two auxiliary pure states are given by~\cite{Sugiura2013canonical}
\begin{equation}
\begin{split}
    |\psi(t)\rangle &= e^{-i\hat{H}t}|\psi\rangle,   \\
    |\phi(t)\rangle &= e^{-i\hat{H}t}\hat{\sigma}_0^z|\psi\rangle.
\end{split}
\label{additional}
\end{equation}
 The reference pure state $|\psi\rangle$ is randomly drawn in Hilbert space, as follows:~\cite{Iitaka2004randomness,Steinigeweg2015typical}
\begin{equation}
    \left|\psi\right\rangle = \sum\limits_{k=1}^{d}(a_k + ib_k)\left|k\right\rangle,
\end{equation}
where the pure state $|k\rangle$ denotes an arbitrary basis of the Hilbert space and the coefficients $a_k$ and $b_k$ are random real numbers, usually drawn from a Gaussian distribution with zero means (other types of randomness are also available)~\cite{Steinigeweg2015typical}. 
Note that the distribution is invariant regardless of any transformations within the Hilbert space. Thus, $\left|\psi\right\rangle$ is a good representative of a thermal statistical ensemble and nearly maximally entangled~\cite{Bartsch2009typical,Vidmar2017maxentanglement,heitmann2020selected}.

In Eq.~\eqref{approximation}, the first term is an approximation of $C_j(t)$, and the second one $\epsilon(|\psi\rangle)$ is the error which is naturally random because of the random choice of typical state $|\psi\rangle$~\cite{Steinigeweg2015typical}. This random error could be eliminated by taking samples over multiple $|\psi\rangle$~\cite{Steinigeweg2015typical}. Notably, the standard deviation of the statistical error $\epsilon = \epsilon(\left|\psi\right\rangle)$ has an upper bound and scales as $\sigma(\epsilon)\propto1/\sqrt{d_{\mathrm{eff}}}$, where $d_{\mathrm{eff}} = \mathrm{Tr}[\mathrm{exp}(-\beta (H-E_0))]$ is the effective dimension of the Hilbert space with $E_0$ being the ground-state energy of $H$~\cite{Bartsch2009typical,Elsayed2013error,Steinigeweg2014error,Steinigeweg2015typical}. Also, $d_{\mathrm{eff}}$ is the number of thermally occupied states and $d_{\mathrm{eff}}=d$ for $\beta =0$~\cite{heitmann2020selected}. Therefore, the error $\epsilon$ exponentially decays with the increase of the system size $L$. 
Remarkably, for our case $\beta\rightarrow0$, DQT is a good approximation~\cite{heitmann2020selected,Steinigeweg2015typical}.

In this way, the trace operation in Eq.~\eqref{full trace} is completely transformed into the time evolution of two auxiliary pure states $|\psi\rangle$ and $\hat{\sigma}_0^z|\psi\rangle$. The full evolution can be subdivided into a product of consecutive steps as~\cite{heitmann2020selected}
\begin{equation}
    |\psi(t)\rangle = \left(e^{-i\hat{H}\delta t}\right)^N |\psi\rangle,
\end{equation}
where each step can be evaluated utilizing a Taylor expansion of the exponential $e^{-i\hat{H}\delta t}$ for sufficiently small $\delta t$ instead of diagonalization. In this paper, we take DQT as a benchmark approach by using an open-source Python package QuSpin~\cite{Phillip2017quspin}. Note that we always consider the entire Hilbert space without any restriction in the context of the DQT method. The number of typical states used in DQT are $5, 2, 1, 1$ for system sizes $L=17, 19, 21, 23$, respectively.

\begin{figure}
\centering
\includegraphics[width=\linewidth]{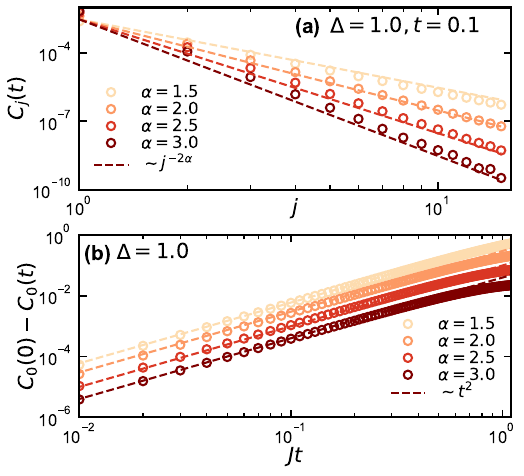}
\caption{
(a) The profile of unequal-time correlation function at early time $t=0.1$ for different $\alpha$ values is well fitted by power-law decay curves $j^{-2\alpha}$~(dashed lines) predicted by the perturbation theory. (b) Short-time dynamics of the autocorrelator for different $\alpha$ values which are fitted by quadratic curves $t^2$~(dashed lines). The numerical results are obtained by the TDVP simulation with a system size of $L=27$.
}
\label{short_time}
\end{figure}

\begin{figure*}
\centering
\includegraphics[width=\linewidth]{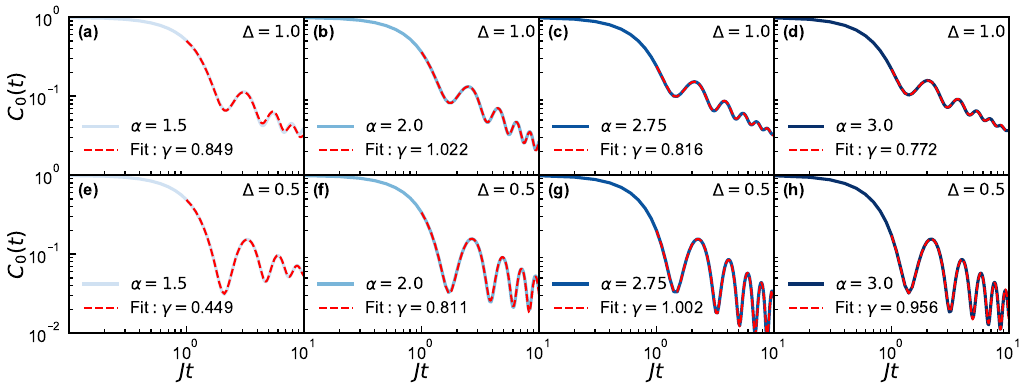}
\caption{
The autocorrelator $C_0(t)$ at $\Delta=1.0$~(a-d) and 0.5~(e-h) for different $\alpha$ values in the log-log scale. The system sizes are all $L=21$ by using the DQT method. The red dashed lines denote the fitting function in Eq.~\eqref{fit}.
}
\label{different delta}
\end{figure*}

\subsection{Time-dependent variational principle}

In addition to the DQT approach, the transport properties can also be investigated utilizing the time-dependent variational principle in the manifold of matrix product states~(TDVP-MPS)~\cite{Haegeman2011tdvp,Haegeman2013tdvp,Haegeman2016tdvp,Kloss2020,Kloss2019longrange}. The matrix product state method allows us to deal with large spin systems, far beyond the state-of-the-art limit of the exact diagonalization (ED) techniques~\cite{Wietek2018ed}.
Moreover, a few macroscopic quantities in the spin chains are explicitly conserved in this method, including the total energy, total magnetization, and total number of particles. For sufficiently large bond dimensions, the method is numerically exact with limitations arising from the growth of entanglement entropy with time, which requires numerical efforts exponentially~\cite{Kim2013limit}. Recent studies have shown that it is possible to handle the time evolution problem of long-range systems using TDVP applied to the manifold of MPS~\cite{Kloss2019longrange,Kloss2020}.

A tensor of order $N$ with finite dimension can be expressed as the product of $N$ matrices, and this representation is essentially a decomposition of tensors. Equivalently, a many-body wave function in Hilbert space
\begin{equation}
    |\Psi\rangle = \sum\limits_{\{i_n\}=1}^d C_{i_1,i_2,\cdots,i_N}|i_1i_2 \cdots i_N\rangle
\label{wave function}
\end{equation}
can be written as a matrix product state
\begin{equation}
    \left|\Psi[A]\right\rangle = \sum\limits_{\{i_n\}=1}^d A^{i_1}(1)A^{i_2}(2)\cdots A^{i_N}(N)|i_1i_2 \cdots i_N\rangle,
\label{MPS}
\end{equation}
where $d$ denotes the local Hilbert space dimension at each site $i_n$ (e.g., $d=2$ in spin-1/2 case), $A^{i_n}(n)$ is a complex matrix with a dimension of $D_{n-1}\times D_{n}$ and the equivalence of Eq.~\eqref{wave function} and Eq.~\eqref{MPS} naturally leads to $D_0=D_L=1$. As the Hilbert space dimension grows exponentially with system size, the dimension (also known as bond dimension) of the matrices $A^{i_n}(n)$ also increases exponentially.
However, for many-body wave functions with low entanglement, we can achieve a good approximation using a series of matrices with small dimensions. This is done by truncating the dimension to a fixed size feasible for computation. This approximation is considered accurate when it converges as we increase this truncated dimension~\cite{Kloss2020}.

Since the typical pure state $|\psi\rangle$ is nearly maximally entangled, it cannot efficiently be represented as an MPS. Instead, we approximate the trace in Eq.~\eqref{full trace} as
\begin{equation}
    C_j(t) \approx \frac{1}{N_\mathrm{s}} \sum\limits_{i=1}^{N_\mathrm{s}} \langle i|\hat{\sigma}_j^z(t)\hat{\sigma}_0^z(0)|i\rangle.
\label{sample}
\end{equation}
To obtain an unbiased average of $C_j(t)$, we sample the product states $|i\rangle$ based on $\sigma^z$ from a probability distribution where each configuration has an equal probability of being chosen~\cite{joshi2022observing}. In practice, it suffices to sample a reasonable number  $N_\mathrm{s}\approx10-100$ of product states to evaluate Eq.~\eqref{full trace}. To further reduce the statistical noise and improve the convergence of Eq.~\eqref{sample}, the conjugate configuration is prepared for each random state, in which all spins except the central one are flipped~\cite{joshi2022observing}.

The main idea of TDVP is to project the time evolution onto the manifold $\mathcal{M}$ of the variational wave function (precisely MPS). This is equivalent to solving the Shr\"{o}dinger equation projected onto the tangent space~\cite{Haegeman2016tdvp,Kloss2020}
\begin{equation}
    i\frac{d|\Psi[A]\rangle}{dt} = P_{\mathcal{M}}\hat{H}|\Psi[A]\rangle,
\end{equation}
where $P_\mathcal{M}$ stands for a projector to the tangent space $\mathcal{M}_\chi$ with fixed bond dimension $\chi$. Since power-law couplings can not be exactly described using the MPS method, we approximate the Hamiltonian by a sum of exponential terms as~\cite{Bull2022XYZ} 
\begin{equation}
    \frac{1}{|i-j|^\alpha} = \sum\limits_n^{N_e} f_n e^{-\lambda_n|i-j|},
\end{equation}
which can be represented efficiently as a matrix product operator (MPO).
By default, we approximate power-law couplings using a sum of $N_e$ exponential functions. These functions are fitted within a range of $L-1$ sites to ensure the resulting approximated couplings deviate from the exact values by less than $2\%$ for any pair of sites~\cite{Kloss2020}.

We use max bond dimension $\chi_\mathrm{max}=512$ (we check its convergence in Appendix~\ref{app:convergence}), sampling number $N_\mathrm{s}=40$, and time step $\delta t=0.01$ as typical parameters in TDVP calculations. Equation~\eqref{sample} is evaluated completely using the TDVP algorithm with the total magnetization conserved in the open-source library TenPy~\cite{Johannes2018tenpy}. In detail, we use a hybrid variation of the TDVP scheme~\cite{Chanda2020MBL,Chanda2020confinement}, where we first use a two-site version of TDVP to dynamically grow the bond dimension to its maximum $\chi_\mathrm{max}$ and then  shift to the one-site version to avoid any errors due to truncation in singular values that appears in the two-site version~\cite{Goto2019tdvp}.

\section{Results and discussions}\label{sec:results}

\subsection{Perturbative short-time dynamics}

At infinite temperature, i.e. $\beta=0$, the trace in Eq.~\eqref{full trace} can be expanded up to the second-order of $t$ at short times. In the Heisenberg picture, we have the expansion of $\sigma_j^z(t)$ as
\begin{equation}
\begin{split}
    \hat{\sigma}_j^z(t) &= e^{i\hat{H}t}\hat{\sigma}_j^ze^{-i\hat{H}t} \\
    &= \hat{\sigma}_j^z + it\left[\hat{H},\hat{\sigma}_j^z\right] + \frac{(it)^2}{2!}\left[\hat{H},\left[\hat{H},\hat{\sigma}_j^z\right]\right] + O(t^3).
\end{split}
\label{expansion}
\end{equation}
After evaluating the commutator, the second term in Eq.~\eqref{expansion} reads 
\begin{equation}
    [\hat{H},\hat{\sigma}_j^z] = -i\sum_{i\neq j}J_{ij}(\hat{\sigma}_i^x\hat{\sigma}_j^y-\hat{\sigma}_i^y\hat{\sigma}_j^x),
\end{equation}
which is $\Delta$-independent and acts as the spin-current operator. The third term of Eq.~\eqref{expansion} has the following complicated form:
\begin{equation}
    \begin{split}
        \left[\hat{H},\left[\hat{H},\hat{\sigma}_j^z\right]\right] &= \sum_{i\neq j}\sum_{k\neq j}J_{ij}J_{kj}(\hat{\sigma}_k^x\hat{\sigma}_i^x\hat{\sigma}_j^z + \hat{\sigma}_k^y\hat{\sigma}_i^y\hat{\sigma}_j^z) \\
        &- \sum_{i\neq k}\sum_{k\neq j}J_{ik}J_{kj}(\hat{\sigma}_i^x\hat{\sigma}_j^x\hat{\sigma}_k^z + \hat{\sigma}_i^y\hat{\sigma}_j^y\hat{\sigma}_k^z) \\
        &- \sum_{i<j}\sum_{k\neq j}J_{ij}J_{kj}\Delta(\hat{\sigma}_i^z\hat{\sigma}_k^x\hat{\sigma}_j^x + \hat{\sigma}_i^z\hat{\sigma}_k^y\hat{\sigma}_j^y) \\
        &+ \sum_{i<k}\sum_{k\neq j}J_{ik}J_{kj}\Delta(\hat{\sigma}_i^z\hat{\sigma}_k^y\hat{\sigma}_j^y + \hat{\sigma}_i^z\hat{\sigma}_k^x\hat{\sigma}_j^x) \\
        &- \sum_{j<i}\sum_{k\neq j}J_{ji}J_{kj}\Delta(\hat{\sigma}_k^x\hat{\sigma}_j^x\hat{\sigma}_i^z + \hat{\sigma}_k^y\hat{\sigma}_j^y\hat{\sigma}_i^z) \\
        &+ \sum_{k<i}\sum_{k\neq j}J_{ki}J_{kj}\Delta(\hat{\sigma}_k^y\hat{\sigma}_j^y\hat{\sigma}_i^z + \hat{\sigma}_k^x\hat{\sigma}_j^x\hat{\sigma}_i^z),
    \end{split}
\end{equation}
which explicitly depends on $\Delta$. Furthermore, considering $\mathrm{Tr}[\hat{\sigma}_i^z\hat{\sigma}_j^z]=\delta_{ij}$ and $\mathrm{Tr}[\cdots\hat{\sigma}_i^{x(y)}\hat{\sigma}_j^{x(y)}\cdots]|_{i\neq j}=0$, the unequal-time correlation function reads
\begin{equation}
C_j(t) \approx \left\{
\begin{aligned}
& 1-t^2\sum_{i\neq j}J_{ij}^2 & & \mathrm{for}\enspace j=0, \\
& t^2J_{j0}^2 & & \mathrm{for}\enspace j\neq 0.
\end{aligned}
\right.
\end{equation}
where $J_{ij}= -\mathcal{N}J/{{\lvert i-j\rvert}^{\alpha}}$ denotes the rescaled hopping strength.

The second-order perturbation theory yields a $\Delta$-independent result. Remarkably, the autocorrelator exhibits quadratic decay early while the spatial correlation function at a fixed time inherits the algebraic decay at lattice distance $j$, falling off as $j^{-2\alpha}$. The short-time dynamics of the correlation function at $\Delta=1.0$ for different $\alpha$ are shown in Fig.~\ref{short_time}, where the spatially algebraic decay Fig.~\ref{short_time}(a) and temporally quadratic decay Fig.~\ref{short_time}(b) of the correlation function are both well captured by fitting lines.

\subsection{Late-time hydrodynamics from autocorrelator}

Beyond the perturbation theory, we now focus on the emergence of late-time hydrodynamic tails, falling off as $t^{-\gamma}$. Expanding the trace in the energy eigenbasis $\hat{H}|n\rangle=E_n|n\rangle$, we have
\begin{equation}
    C_j(t) = \sum_{n,m}\langle n|\hat{\sigma}_j^z|m\rangle\langle m|\hat{\sigma}_0^z|n\rangle e^{-i(E_n-E_m)t}.
\end{equation}
The autocorrelator $C_0(t)$ consists of a superposition of $\cos\left[(E_n-E_m)t\right]$. The weight of each cosine function is determined by the matrix elements of $\hat{\sigma}^z$ in the energy basis. Such a form is reminiscent of the spectral form factor defined by $K(t)=\sum_{n,m}e^{-i(E_n-E_m)t}$. 
One may expect that $C_0(t)$ exhibits oscillations early, as seen in other spin models~\cite{Richter2019xxz}.

The autocorrelator $C_0(t)$ at $\Delta=1.0$ and 0.5 for different $\alpha$ are demonstrated in Fig.~\ref{different delta}. 
Early, typically $Jt<1$, the autocorrelator presents rapid decay so that it reaches a local equilibrium. Afterward, the autocorrelator suddenly increases and exhibits slow hydrodynamic transport constrained by the conserved magnetization at an intermediate time. Eventually, the system enters a global equilibrium. More evidence of these multiple stages can also be found in the inset of Fig.~\ref{autocorr}~(see in Appendix~\ref{app:comparison}).

As expected, the power-law decay exists but is hindered by oscillations at late time~($t>1$). This is similar to the imbalance dynamics in the presence of disorder. To extract the dynamical exponent $\gamma$, we find a proper fitting function to capture the relaxation at late time~($t>1$)~\cite{Luitz2016MBL}
\begin{equation}
    C_0(t) = ae^{-t/\tau}\mathrm{cos}(\beta t+\phi) + bt^{-\gamma}[1+ct^{-\eta}\mathrm{sin}(\omega t+\phi)].
\label{fit}
\end{equation}
The second term has a primary contribution and it consists of a primary algebraic tail $t^{-\gamma}$ and damped oscillation with specific frequency $\omega$. 
For $\Delta=1.0$, as $\alpha$ increases from $1.5$ up to $3.0$, we find that the dynamical exponent firstly increases until it reaches its maximum value of $\gamma\simeq 1$ at $\alpha=2.0$, and then it dramatically decreases (see Fig.~\ref{different delta}(a-d)). 
Since $\gamma$ is always higher than $1/2$, this non-monotonic behavior of $\gamma$ could be considered as a superdiffusive-ballistic-superdiffusive transition.
Similar non-monotonic behavior also exists at $\Delta=0.5$ as shown in Fig.~\ref{different delta}(e-h). What differs is that ballistic transport occurs at $\alpha=2.75$ for $\Delta=0.5$, indicating that this transition boundary is highly related to both $\Delta$ and $\alpha$.
Also, we notice that subdiffusion ($\gamma<1/2$) appears at $\alpha=1.5$. 

Then, based on the fitting from Eq.~\eqref{fit}, we have the phase diagram of transport properties, i.e. the dynamical exponent $\gamma$ as a function of $\Delta$ and $\alpha$ as shown in Fig.~\ref{phase}(a). There are three kinds of phase: subdiffusion~($\gamma<1/2$), superdiffusion~($1/2<\gamma<1$) and ballistic transport~($\gamma\simeq1$). 
Remarkably, an anomalous superdiffusive-ballistic-superdiffusive transition generally exists at an arbitrary anisotropy $\Delta$. We find the peak value of such transitions can be greatly captured by an exponential fitting curve $\Delta\sim e^{-\alpha+2}$ (red dashed curve).
This anomalous hydrodynamics strongly violates the expected diffusion in a non-integrable system, which is one of the highlights of our work.

Intuitively, such a non-monotonic behavior of $\gamma$ is induced by anisotropy $\Delta$ against long-range exponent $\alpha$. It has been shown that strong anisotropy acts as an attractive potential to bound two spins with parallel polarization in ferromagnetic phase~($\Delta>0$), leading to bound states at large $\Delta$~\cite{Macr2021bound,Kranzl2023bound}. Thus, the anisotropy can slow down the spreading of spins. On the contrary, long-range couplings tend to rapidly relax spins away from initial excitation. Then ballistic transport emerges when the effects of them are balanced. Since ballistic transport generally exists in integrable systems, our model at the transition point might be linked to some integrable models. In particular, we note that at $\Delta=1.0$ and $\alpha=2.0$, Eq.~\eqref{hamiltonian} in the thermodynamic limit corresponds to the well-known Haldane-Shastry integrable model~\cite{Haldane1988,shastry1988exact}. 

\begin{figure}
\centering
\includegraphics[width=\linewidth]{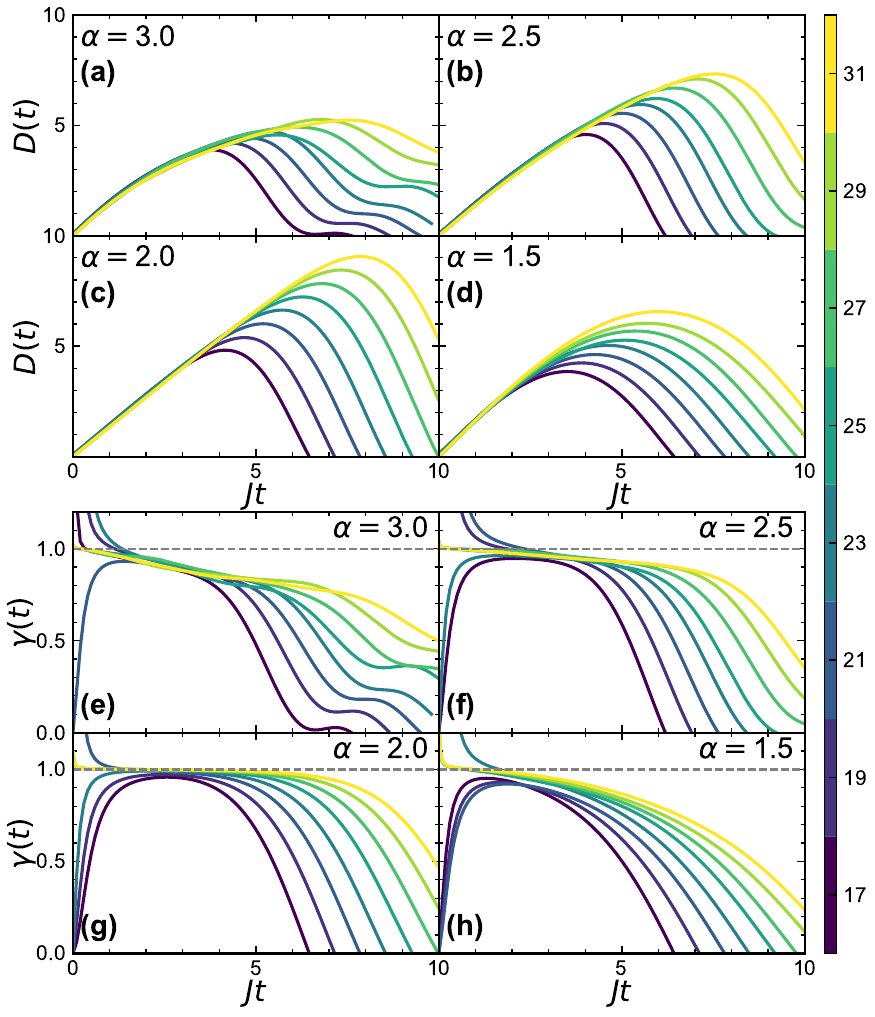}
\caption{
Diffusion coefficient $D(t)$ for $\alpha=3.0$~(a), $2.5$~(b), $2.0$~(c), $1.5$~(d) with different system sizes. Dynamical exponent $\gamma(t)$ for $\alpha=3.0$~(e), $2.5$~(f), $2.0$~(g), $1.5$~(h) with different system sizes. We utilize the DQT method for $L=17, 19, 21, 23$ and the TDVP technique for $L=25, 27, 29, 31$.
}
\label{diffusive_scale}
\end{figure}

\subsection{Emergent region against quantum chaos}
Typically, a quantum integrable system possesses an extensive set of conserved quantities, which prevent the system from quantum chaos and its information loss.
These conserved quantities represent observables whose expectation values remain constant as the system evolves. As shown in the following section, the spin current operator $\hat{J}$ exhibits near-constant behavior during ballistic transport, indicating a corresponding conservation law in the Hamiltonian.
Utilizing the random matrix theory (RMT), we find that the energy level-spacing statistics of our Hamiltonian in the same parameter space $(\Delta,\alpha)$ in Fig.~\ref{phase}(a) mostly obey the Wigner-Dyson (WD) distribution with a ratio $\langle r\rangle\approx 0.53$ (see in Fig.~\ref{r_value}(b) and (d)). This indicates that this long-range spin system is non-integrable, contradicting our previous understanding.

To further quantify the quantum chaos beyond the RMT, we investigate the microcanonical distributions of eigenstate entanglement entropy (EE) by following recent Ref.~\cite{Joaquin2023quantifying}. For a chosen eigenstate $\lvert \psi\rangle$, the von Neumann entanglement entropy of a subsystem $A$ is defined as $S_A=-\mathrm{Tr}[\rho_A\mathrm{log}\rho_A]$ with the reduced density matrix $\rho_A=\mathrm{Tr}_B[\lvert \psi\rangle\langle\psi\rvert]$. For a Hamiltonian system, the central idea amounts to comparing the subsystem's EE distribution $P_{\mathrm{E}}(S_A)$ of eigenstates at the mid spectrum to the reference distribution of a pure random state, i.e., the Bianchi-Dona (BD) distribution $P_{\mathrm{BD}}(S_A)$. The distance between distributions is characterized by the Kullback-Leibler (KL) divergence $D_{\mathrm{KL}}$. If we assume that $P_{\mathrm{E}}(S_A)$ and $P_{\mathrm{BD}}(S_A)$ are Gaussian distributions, the KL divergence is given by
\begin{equation}
    D_{\mathrm{KL}}=\frac{(\mu_{\mathrm{E}}-\mu_{\mathrm{BD}})^2}{2\sigma^2_{\mathrm{BD}}}+\frac{1}{2}
    \left[\left(\frac{\sigma_{\mathrm{E}}}{\sigma_{\mathrm{BD}}}\right)^2-1\right]-\mathrm{log}\frac{\sigma_{\mathrm{E}}}{\sigma_{\mathrm{BD}}},
\end{equation}
where $\mu$ and $\sigma$ are the mean value and the standard deviation of EE distribution $P(S_A)$, respectively. 

Then, we choose the typical system sizes as $L=12$ and $14$ with the subsystem size $L_A=L/2$ in the half-filling sector $\sum_{i}\hat{\sigma}_{i}^z=0$. For $L=12$, the mean and standard deviation of the BD distribution is $\mu_\mathrm{BD}\approx 3.5745$ and $\sigma_\mathrm{BD}\approx 0.0199$, respectively. For $L=14$, $\mu_\mathrm{BD}\approx 4.2652$ and $\sigma_\mathrm{BD}\approx 0.0103$~\cite{Joaquin2023quantifying}.
We select $100$ eigenstates ($L=12$) and $300$ eigenstates ($L=14$), lying in a small energy window whose density of states (DOS) is the largest to compute the half-chain EE and the corresponding $\mu_\mathrm{E}$ and $\sigma_\mathrm{E}$. To further remove the fluctuations from sampling eigenstates, we add a weak disorder $h_i\sum_i \hat{\sigma}_i^z$ to Eq.~\eqref{hamiltonian} with $h_i$ randomly drawn from $[-0.1J, 0.1J]$. The EE distribution is averaged over $100$ disorder realizations for both $L=12$ and $14$, respectively. The divergence $D_{\mathrm{KL}}$ as a function of $(\Delta,\alpha)$ is plotted in Fig.~\ref{phase}(b). Since our model at $\alpha\rightarrow\infty$ reduces to the integrable Heisenberg model, the divergence increases monotonically with increasing $\alpha$. 
Surprisingly, for different values of $\Delta$, a narrow region consistently emerges (shown in dark purple) where $D_{\mathrm{KL}}$ reaches a local maximum after resuming its monotonic rise (see also in Appendix~\ref{app:KL and r}). These unexpected local maxima defy the typical behavior associated with quantum chaos, even though the model is considered ergodic. Remarkably, the points $(\Delta,\alpha)$ at which these maxima occur can all be described by the same exponential curve $\Delta\simeq e^{-\alpha+2}$~(red dashed curve) in Fig.~\ref{phase}(a). This suggests that our model exhibits an emergent property that resists quantum chaos at specific points $(\Delta,\alpha)$ around this exponential fitting curve. This finding aligns with the anomalous ballistic transport in Fig.~\ref{phase}(a), which is the other highlight of our work.

\subsection{Broadening of spin excitation}

\begin{figure}
\centering
\includegraphics[width=\linewidth]{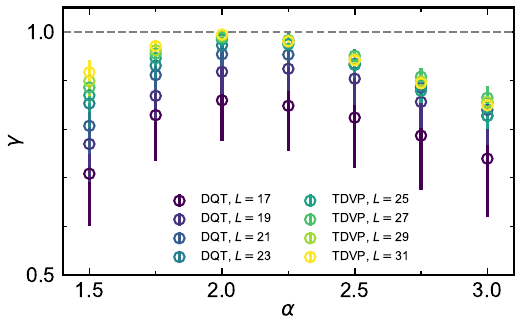}
\caption{
Dynamical exponent $\gamma$ as a function of $\alpha$ for different system sizes. Dynamical exponent $\gamma$ is obtained by averaging $\gamma(t)$ over the time window of $Jt\in[3, 5]$. The error bars denote the standard deviation of $\gamma$.
The dashed line denotes the ballistic transport ($\gamma=1$). 
}
\label{dynamical_exponent}
\end{figure}

To further confirm this anomalous phenomenon, we now investigate the mean-square displacement~(MSD). This quantity reflects the spread of spin excitation over space and time. Here, we focus on the isotropic case $\Delta=1.0$ to quantitatively investigate the non-monotonic behavior of $\gamma$ with varying $\alpha$. We can extract the diffusion coefficient $D(t)\propto t^{2\gamma-1}$ and dynamical exponent $\gamma(t)$ from Eq.~\eqref{MSD}. 
Diffusion coefficients for different values of $\alpha$ with different system sizes are shown in Fig.~\ref{diffusive_scale}(a-d). Note that the results of DQT and TDVP methods agree with each other for all $\alpha$ in Fig.~\ref{diffusive_scale}(a-d). 
Increasing the system size does not halt the growth of $D(t)$ for all the considered values of  $\alpha$. This excludes diffusive transport since a diffusion process would exhibit a size-independent plateau (constant $D(t)$) at large system sizes.  Moreover, we find that $D(t)$  grows rapidly and nearly linearly at $\alpha = 2.0$, then slows down at $\alpha = 1.5$. This non-monotonic behavior of $D(t)$  reflects the transition again from superdiffusive to ballistic and then back to superdiffusive transport.

In Fig.~\ref{diffusive_scale}(e-h), we extract the time-dependent dynamical exponents $\gamma(t)$ for various values of $\alpha$ and system sizes. This allows us to quantitatively distinguish between different transports. In the ideal scenario, $\gamma(t)$ would converge at a finite value as time approaches infinity $\lim\limits_{t\to\infty}\gamma(t)\approx\gamma$. For a finite system, the plateau in $\gamma(t)$ is short-lived.
However, as the system size increases, $\gamma(t)$ gradually converges to a stable curve for all the considered values of $\alpha$.
We also find that the dynamical exponent in Fig.~\ref{diffusive_scale}(e-h) is always larger than $1/2$, indicating the absence of diffusion. Remarkably, $\gamma(t)$ is quite close to $1$ at $\alpha=2.0$ where we believe ballistic transport occurs. The ballistic transport at $\alpha=2.0$ is further verified by the spatial correlation function in Fig.~\ref{curvefit}~(see in Appendix~\ref{app:comparison}). 
Interestingly, the density profile at the central part of the chain is different from the Gaussian fitting curves (dash curves in Fig.~\ref{curvefit}). This indicates that the transport mechanism is beyond the diffusion.
The values of $\gamma(t)$ for other $\alpha$ are smaller than  $1$ and this implies a non-monotonic change in transport behavior.
Figure~\ref{dynamical_exponent} summarizes the dynamical exponent $\gamma$ as a function of some typical values of $\alpha$ for different system sizes. The stable dynamical exponents, represented by $\gamma$ are obtained by averaging over the stable plateaus in Fig.~\ref{diffusive_scale}(e-h). As expected, the non-monotonic behavior of $\gamma$ persists for various system sizes, with ballistic transport occurring around $\alpha = 2.0$.

\subsection{Understanding from the Kubo formula}

\begin{figure}
\centering
\includegraphics[width=\linewidth]{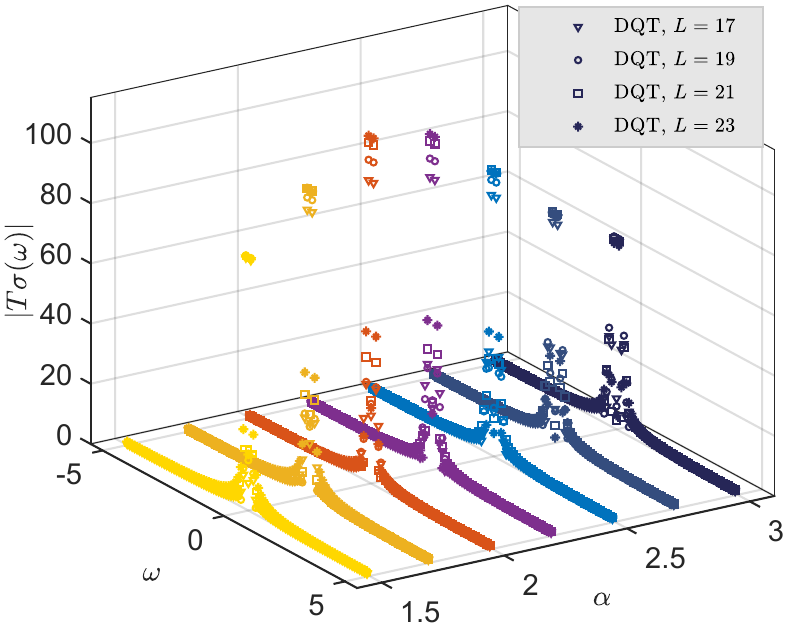}
\caption{
The spin conductivity $|T\sigma(\omega)|$ at $\Delta=1.0$ for different $\alpha$ values and different system sizes.
}
\label{conductivity}
\end{figure}
Linear response theory provides a universal framework for understanding a system's response to an external force treated as a perturbation. It allows us to understand this superdiffusive-ballistic-superdiffusive transition in the frequency domain. Thus, the transport behavior can be characterized by the spin conductivity $\sigma(\omega)$, which reads~\cite{Bertini2021review,gopalakrishnan2022anomalous}
 \begin{equation}
     T\sigma(\omega) = \frac{1}{L} \int_{0}^{\infty} dt \left\langle \hat{J}(t)\hat{J}(0)\right\rangle e^{i\omega t}, 
 \label{kubo}    
 \end{equation}
where the current operator $\hat{J}=\sum_{l} \hat{J}_l$ is defined by the lattice continuity equation
$\partial_{t}\hat{\sigma}_l^z=i[\hat{H},\hat{\sigma}_l^z]=\hat{J}_{l-1}-\hat{J}_l$. Note that $T\sigma(\omega)$ has a finite value since $\sigma(\omega)\rightarrow0$ at infinite temperature limit $T\rightarrow\infty$, implying that the system does not respond to any perturbations. Its continuous form leads to a generalized relation as~\cite{Jacopo2019kubo,Singh2023kubo}
\begin{equation}
    \frac{1}{2}\frac{d^2}{dt^2}\Sigma^2(t)=\frac{1}{L}\left\langle \hat{J}(t)\hat{J}(0)\right\rangle.
\label{relation}
\end{equation}
Then, substituting Eq.~\eqref{relation} into Eq.~\eqref{kubo} and taking $\Sigma^2(t)\sim t^{2\gamma}$, we have
\begin{equation}
    T\sigma'(\omega)\sim \omega^{1-2\gamma},
\end{equation}
where $\sigma'(\omega)$ is the real part of $\sigma(\omega)$. This means that the static response diverges when transports are ballistic or superdiffusive~(i.e. $1/2<\gamma\leq1$). Then we can decompose $\sigma'(\omega)$ into a singular and a regular part~\cite{Bertini2021review}
\begin{equation}
    T\sigma'(\omega) = TD_\mathrm{Drude}\delta(\omega) + T\sigma_\mathrm{reg}(\omega)
\end{equation}
where the $D_\mathrm{Drude}$ is the so-called Drude weight. Since Drude weight trivially vanishes at $T\rightarrow\infty$, one can expect $TD_\mathrm{Drude}$ to be finite.
Referring back to Eq.~\eqref{kubo}, this result signifies that the injected currents do not completely decay, i.e., ballistic transport in an ideal conductor. 
Notably, for systems with a finite local Hilbert space, the singularity in the conductivity cannot be stronger than  $\omega^{-1}$~\cite{Bertini2021review}. 
Consequently, a non-zero Drude conductivity ($TD_\mathrm{Drude}\neq0$) only occurs when the ballistic transport is present (i.e. $\gamma = 1$). In this case, this $\omega^{-1}$ term is included in the $\delta$-function of the Kubo formula, while weaker singularities are retained in the part of $\sigma_\mathrm{reg}(\omega)$.

Here we focus on the divergence of $|T\sigma(\omega=0)|$ instead of giving the exact value of $TD_\mathrm{Drude}$. We show the absolute value of conductivity for typical $\alpha$ values and different system sizes in Fig.~\ref{conductivity}, where $T\sigma(\omega)$ is obtained from the Fourier transform of Eq.~\eqref{relation}. As expected, all conductivities become increasingly peaked around $\omega=0$, indicating that $\gamma>1/2$. Moreover, as $\alpha$ increases, the peak becomes higher and it reaches a maximum value at $\alpha=2.0$ followed by a decrease. This non-monotonic behavior generally exists independent of system sizes. $T\sigma(0)$ at $\alpha=2.0$ has the largest singularity, indicating the existence of ballistic transport, and agrees with the MSD analysis above. One can also find that, for $\gamma=1$, the current-current correlator $\langle \hat{J}(t)\hat{J}(0) \rangle$ is a constant, implying that the current operator $\hat{J}$ commutes with the Hamiltonian. This is manifested on the emergent local maximum of $D_\mathrm{KL}$.

\section{Conclusion}\label{sec:conclusion}

In summary, we have employed both DQT and TDVP approaches to numerically study the unequal-time spin-spin correlations at infinite temperature in a long-range spin-$1/2$ XXZ model. By varying the long-range exponent $\alpha$, we find an anomalous superdiffusive-ballistic-superdiffusive transition for all anisotropy values in $\Delta\in( 0, 1.5]$.
This phenomenon is also supported by the static spin conductivity $T\sigma(\omega=0)$, which exhibits the largest singularity in the presence of ballistic transport. Moreover, the boundary of this transition can be well fitted by an exponential curve $\Delta\simeq e^{-\alpha+2}$. 
This transition is caused by the interplay between anisotropy and long-range couplings.
We further investigate the Kullback-Leibler divergence between the entanglement entropy's microcanonical distribution of eigenstates and that of a purely random state. This analysis suggests that the ergodic long-range XXZ model exhibits resistance to quantum chaos at the transition boundary. 
Interestingly, similar signatures could be present in anomalous hydrodynamic behaviors in other non-integrable quantum many-body systems. 
Since these transports occur at high temperatures, it would be useful to study the classical description of spin density relaxation in the long-range XXZ model. However, understanding the mechanism behind this ballistic transport and its connection to the emergent resistance to quantum chaos in non-integrable quantum systems remains an open question.

We study the magnetization dynamics above in a restricted range of anisotropy $\Delta$ and long-range exponent $\alpha$. By expanding these parameter ranges, the system can exhibit various types of transport behavior for spin relaxation, including subdiffusive, diffusive, superdiffusive, and ballistic. We expect there to be optimal points where standard diffusion and KPZ superdiffusion become dominant. 
While preparing this manuscript, we discovered an independent study by M. Mierzejewski et al.~\cite{Mierzejewski2023quasi}. Their work suggests 
the presence of quasi-ballistic transport in a similar model. However, their approach using a single domain-wall state makes it hard to capture the full picture of infinite-temperature spin transport. In addition, their conductivity analysis is limited to the specific case where the total magnetization is zero, i.e. $\sum_{i}\hat{S}_{i}^z=0$.
On the contrary, our work focuses on the infinite-temperature limit, involving the entire Hilbert space. This approach also allows us to measure the autocorrelator by averaging over product states, which aligns with how hydrodynamics are observed in experiments. Although recent work has utilized quantum circuits to directly measure the infinite-temperature autocorrelator without needing to sample product states~\cite{shi2023probing}, our study offers the advantage of being readily adaptable to finite temperature regimes, where spin relaxation might be slower.

\section*{Acknowledgment}
We thank Dr. Dario Poletti for helpful discussions.
This work was supported by National Natural Science Foundation of China under Grants No.~12375021 and National Key R\&D Program of China under grants No. 2022YFA1404203.


\appendix
\section{Convergence of TDVP results with bond dimension and random state samples}\label{app:convergence}
\begin{figure}
\centering
\includegraphics[width=\linewidth]{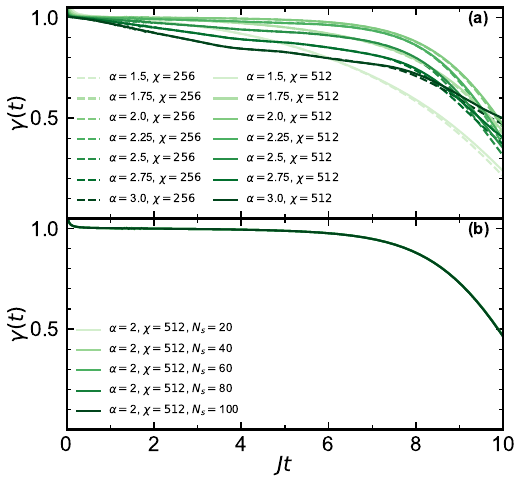}
\caption{
(a) Time-dependent dynamical exponent $\gamma(t)$ for different $\alpha$ values and bond dimensions $\chi$. The number of random state samples is $N_\mathrm{s}=40$. (b) Time-dependent dynamical exponent $\gamma(t)$ at $\alpha=2.0$ for different samples $N=20,40,60,80,100$, respectively. The bond dimension is $\chi=512$ and the system size in both (a) and (b) is 
$L=31$ for TDVP simulations.}
\label{convergence}
\end{figure}

Here we briefly show that the dynamical exponent $\gamma(t)$ obtained by the TDVP method is converged with bond dimension and samples. We focus on the isotropic case $\Delta=1.0$. As shown in Fig.~\ref{convergence}(a), despite different $\alpha$, $\gamma(t)$ are all converged for a long time ($Jt\approx8$) with increasing bond dimension. The flat part of $\gamma(t)$ is reliable for extracting a stable dynamical exponent $\gamma$. Besides, ballistic transport is expected to occur at $\alpha=2.0$, hence we try different sampling numbers of random states to check the result. With fixed $\chi$ and increasing samples up to $N=100$ in Fig.~\ref{convergence}(b), $\gamma(t)$ perfectly converges. We have checked that $\gamma(t)$ for other $\alpha$ is also converged with sampling numbers, and $N_\mathrm{s}=40$ is enough for TDVP calculation. 

\section{Comparison of DQT and TDVP results}\label{app:comparison}

\begin{figure}
\centering
\includegraphics[width=\linewidth]{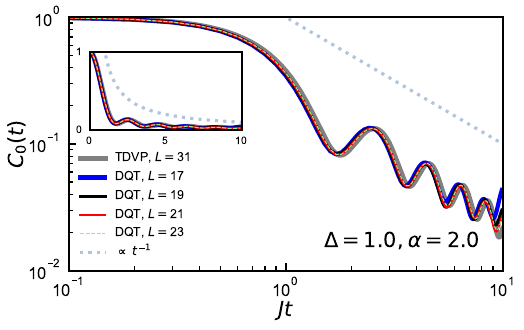}
\caption{
Autocorrelators $C_0(t)$ for $\alpha=2.0$ and $\Delta=1.0$ with system sizes $L=17, 19, 21, 23$~(DQT), $31$~(TDVP) in the log-log scale. The lightsteelbule dashed line indicates ballistic transport $\propto t^{-1}$. The inset shows the same quantities but on a linear scale.
}
\label{autocorr}
\end{figure}

\begin{figure*}
\centering
\includegraphics[width=\linewidth]{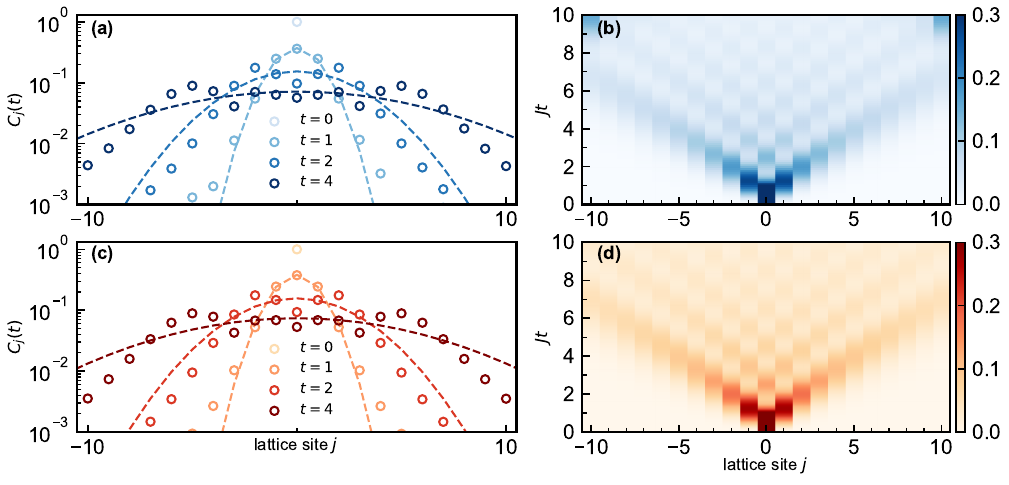}
\caption{
(a,b) Density profiles for $\alpha=2.0$ and $\Delta=1.0$ with $L=23$~(DQT) (a) and the corresponding broadening of spin excitation (b). (c,d) The same data with $L=31$~(TDVP). Dashed curves are Gaussian fits to the data in (a)(c).
}
\label{curvefit}
\end{figure*}

\begin{figure*}
\centering
\includegraphics[width=0.9\linewidth]{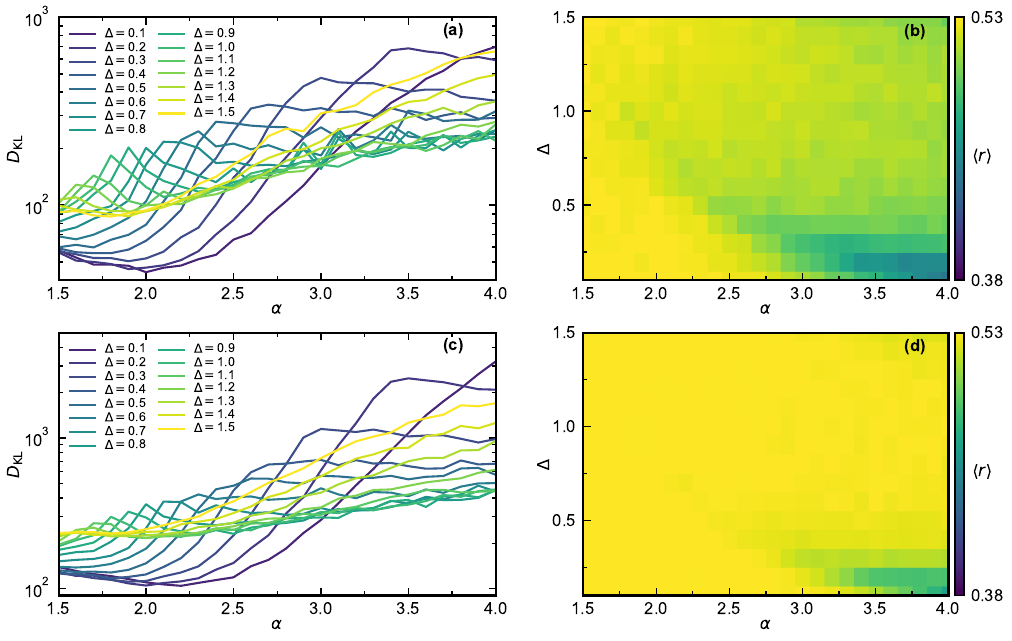}
\caption{
(a-b) The Kullback-Leibler divergence (a) and the level-spacing ratio (b) for system size $L=12$ at the half-filling condition. (c-d) The same data with (a-b) for system size $L=14$ at the half-filling condition.
}
\label{r_value}
\end{figure*}

To verify the accuracy of the TDVP method, we briefly compare the autocorrelator and broadening of spin density obtained from DQT and TDVP methods. Figure~\ref{autocorr} shows the autocorrelators with different system sizes at $\alpha=2.0$ and $\Delta=1.0$ where ballistic transport is present. One finds that as the system size increases, different autocorrelators gradually converge to a stable curve. This curve roughly matches the ballistic decay. Besides, benefiting from the large system size, the TDVP method efficiently improves the boundary effect (see Fig.~\ref{curvefit}(b-d)). The density profiles obtained from the two methods are also in good agreement at the intermediate time scale.

\section{Level statistics and the KL divergence}\label{app:KL and r}

By using exact diagonalization, we compute the consecutive energy gaps $\delta_k=E_{k+1}-E_k$ of the Hamiltonian. The level statistics are characterized by their ratio $r_k=\mathrm{min}(\delta_{k+1},\delta{k})/\mathrm{max}(\delta_{k+1},\delta{k})$. The averaged value $\langle r\rangle$ allows one to distinguish between the Poisson and Wigner-Dyson level statistics. The ergodic phase is known to follow the Wigner-Dyson distribution with $\langle r\rangle\approx 0.53$. In Fig.~\ref{r_value}(b) and (d), we illustrate the ratios $\langle r\rangle$ for all $(\Delta,\alpha)$ considered here, which is mostly fluctuated around $0.53$, indicating that the system is ergodic.

The Kullback-Leibler (KL) divergence between distributions $P_{\mathrm{E}}(S_A)$ and $P_{\mathrm{BD}}(S_A)$ is defined as
\begin{equation}
    D_{\mathrm{KL}}(P_{\mathrm{E}},P_{\mathrm{BD}})=\int dS_A P_{\mathrm{E}}(S_A)\mathrm{log}\frac{P_{\mathrm{E}}(S_A)}{P_{\mathrm{BD}}(S_A)}.
\end{equation}
The local maximum of $D_\mathrm{KL}$ for different $\Delta$ is shown in Fig.~\ref{r_value}(a) and (c). One finds that the emergent local maximums are robust with increasing system size.


%

\end{document}